\documentclass[preprint2]{aastex}
\usepackage{natbib}
\shorttitle{Easylife for VIPERS}
\shortauthors{Garilli et al.}

\begin{document}

\title{Easylife: the data reduction and survey handling system for
VIPERS} 
\author{B. Garilli, L. Paioro, M. Scodeggio, P. Franzetti, M. Fumana}
\affil{INAF-IASF Milano, via Bassini 15, 20133 - Milano (Italy)}
\author{L. Guzzo}
\affil{INAF-Osservatorio Astronomico di Brera, Via Bianchi 46, 23807 -
  Merate, Italy}

\begin{abstract}
We present Easylife, the software environment developed within the framework of the VIPERS
project for  automatic data
reduction and survey handling. Easylife is a comprehensive system to automatically reduce
spectroscopic data, to monitor the survey advancement at all stages, to
distribute data within the collaboration and to release data to the whole community. It is based on
the OPTICON founded project FASE, and inherits the FASE capabilities of modularity
and scalability. After describing the software architecture, the main reduction and quality
control features and the main services made available, we show its performance in terms of 
reliability of results. 
We also show how it can be ported to other projects having different characteristics.

\end{abstract}


\section{Introduction}

Thanks to the continuous evolution of astronomical instrumentation and
in particular of the
multiplexing gain of faint-object spectrographs, large-scale
spectroscopic surveys have become a real industry, in which up
to $10^6$ spectra can be accumulated 
by a single project.  So far, this has been particularly true for redshift surveys
of the ``local'' Universe ($z\sim 0.1$), with the notable milestones
represented by the 2dF
Galaxy Redshift Survey (2dFGRS, \citealt{2df}) and the Sloan Digital Sky
Survey (SDSS, \citealt{sdss1,sdss7}), which built over earlier pioneering projects
of the 1980's and 1990's
as e.g. CfA redshift survey \citep{Davis1980CfA1,Geller1988CfA2}, Perseus-Pisces
\citep{GiovanelliHaynes1986}, ESP \citep{ESP} and LCRS
\citep{LasCampanas}.  

For obvius reasons, redshift surveys of the more distant Universe ($z\sim 1$), 
were limited to smaller numbers, with samples of a few hundred to a
few thousands galaxies 
(e.g. CFRS, \citealt{CFRS}), which more recently grew to a few tens
of thousands objects, with the advent of new multi-object spectrographs on 8-m class
telescopes, like VIMOS and DEIMOS
(e.g. VVDS, \citealt{vvds_main,vvds_wide}; 
DEEP2, \citealt{deep2}; zCosmos, \citealt{zcosmos_main}).  Lately, a
further increase in the size of samples at intermediate redshift
($z\sim 0.5$) has been possible by targetting specific classes of
galaxies, like star-forming objects in the case of the WiggleZ survey
\citep{WiggleZ} or massive ``reddish'' galaxies in the case of
SDSS3-BOSS \citep{BOSS}.  In particular, the total yield for this latter project 
will be of the order of  $10^6$  spectra.  This trend is expected to
continue with future redshift surveys, as it is the case for the tens of millions
redshifts expected for the ESA mission Euclid \citep{euclidRB}, or Gaia \citep{gaia1,gaia2}.

Potentially, the amount of information provided by such large-scale
surveys is enormous, but to exploit its full scientific potential,
measurements have to be extracted from the raw data in a way which is both
efficient (thus with minimum human intervention) and at the same time reliable;
equally important, they have to be distributed to the
community in an easily manageable form. With these goals in mind, a
number of projects have developed 
automatic pipelines tuned for their own needs. 

Since its planning
in the early 1990s, the SDSS dedicated a big effort to the creation of
a full pipeline for the data reduction (e.g. \citealt{Lupton2002}) and a parallel database system
to handle the enormous (for that time) amount of photometric and
spectroscopic data (e.g. \citealt{Szalay2002}).  Similar efforts were later implemented in
particular by large photometric surveys, with the creation of
data processing centres, like {\it Terapix} for the CFHT
observations \citep{terapix}, the UKIDSS center \citep{ukidss} 
or the CANDELS pipeline \citep{candels}.  Among pure spectroscopic surveys, VVDS
\citep{vipgi}, zCosmos \citep{zcosmos_main} AGES
\citep{ages} and more recently WiggleZ \citep{WiggleZ} have all built their
own tools, eventually glueing together  pre-existing algorithms and programs into an
automatic processing.

Data dissemination is the second important requirement these projects
have to face. This includes both internal distribution to the survey
team, and public release to the scientific
community; the latter may also include public outreach products.  The Virtual Observatory
has set up standards and conventional formats for this purpose (see
http://www.ivoa.net/Documents/).  VO compatible tools have been 
flourishing over recent years (Aladin, Topcat, VOSpec: for a more complete list see
http://www.ivoa.net/cgi\--bin/twi\-ki/bin/vi\-ew/IV\-OA/Iv\-oa\-Appli\-cations),
and are on the way of becoming the standard for data dissemination.
Currently, however, each survey still tends to provide its own
specific web pages, from where data and information can be downloaded,
either through plain ASCII files or via more sophisticated
database systems: SDSS (http://www.sd\-ss.org/dr\-7/), DEEP
(http://deep.berke\-ley.e\-du/DR3/), VVDS
(ht\-tp://ce\-sam.oa\-mp.fr/vv\-ds\-project/) 
and Cosmos (http://cos\-mos.astro.cal\-tech.e\-du/da\-ta), are just some
examples.

A third important point in exploiting such large and long-lasting projects is
book-keeping of the survey processes. This is usually kept by
the project coordinator or by a restricted coordination group, not
always using appropriate tools, with considerable expenditure of time. 

When we started  the VIMOS Public Extragalactic Redshift
Survey (VIPERS) in 2008, we decided to invest time and manpower
in a survey management system capable of automatically taking care 
of: data
reduction and redshift measurement, quality control, data dissemination (both
internal and to the public) and logging. In this paper we describe the system
we have set up, called {\it Easylife}. In section \ref{vipers},\ref{vimosInstr} and \ref{vimos}
we briefly describe the VIMOS Public Extragalactic Redshift Survey (VIPERS \footnote{\url{http://vipers.inaf.it}}) survey, the VIMOS spectrograph and the observing sequence to be followed
within ESO projects. In section \ref{requirements} we detail the requirements we
  have defined. The system architecture is briefly outlined in section
  \ref{architecture}. After a description of the main tools (section \ref{blocks}), we
  dedicate section \ref{performances} to the performance in terms of
  reduction quality we obtain with {\it Easylife}. In section \ref{LBT} we show  how
  we are using  {\it Easylife} for other projects.

\section{The VIPERS survey}
\label{vipers}

VIPERS is an ongoing ESO Large
Programme aimed at measuring redshifts for $\sim 10^5$ galaxies at
redshift $0.5 < z \lesssim 1.2$, to accurately and robustly measure clustering, the
growth of structure (through redshift-space distortions) and galaxy
properties at an epoch when the Universe was about half its current
age.  
The galaxy sample is selected from the 
Canada-France-Hawaii Telescope Legacy Survey Wide (CFHTLS-Wide)
optical photometric catalogues \citep{CFHTLS}.  VIPERS covers
$\sim24$ deg$^2$ on the sky, divided over two areas within the W1 and W4 CFHTLS
fields.  Galaxies are selected to a limit of $i_{AB}<22.5$, further
applying a simple and robust $gri$ colour
pre-selection, as to effectively remove
galaxies at $z<0.5$. Coupled to an aggressive observing strategy
\citep{shortSlit}, this allows us to double the galaxy sampling
rate in the redshift
range of interest, with respect to a pure magnitude-limited sample,
reaching a target sampling rate sampling of $\sim 40\%$.  At the same time, the area and depth of the survey 
result in a fairly large volume, $5 \times 10^{7}$ h$^{-3}$ Mpc$^{3}$, analogous
to that of the 2dFGRS at $z\sim0.1$.  Such combination of  sampling and
depth is quite unique over current redshift surveys at $z>0.5$.
VIPERS Spectra are collected with the VIMOS multi-object
spectrograph \citep{instrument1} at moderate resolution ($R=210$), using the LR Red
grism, providing a wavelength coverage of 5500-9500${\AA}$ and a
typical radial velocity error of $175(1+z)$ km sec$^{-1}$ .  The full
VIPERS area of $\sim 24$ deg$^2$ is covered through a mosaic 
of 288 VIMOS pointings (192 in the W1 area, and 96 in the W4 area).  

As of January 2012, about 60\% of the VIPERS area has been
observed, with completion expected by 2014.  A first discussion of the
spectral data together with Principal Component classification can be found in
\citet{Marchetti2012}.  More details will be available in Guzzo et al.
(2012, in preparation).   

\section{The VIMOS spectrograph}
\label{vimosInstr}
VIMOS (VIsible MultiObject Spectrograph) is an imaging spectrograph installed on Unit 3 (Melipal) 
of the ESO Very Large Telescope (VLT) at the Paranal Observatory 
in Chile (see \citet{instrument1} and \citet{instrument2} for a detailed 
description of the instrument and its capabilities). 
The driving design concept for the instrument is to cover 
as much of the unvignetted part of the focal plane as possible at 
the VLT Nasmyth focus (a circular area with a diameter of 22 arcmin on the sky). 
Since this large area corresponds to a very large linear scale (almost 1 m), 
it was decided that coverage would be achieved by splitting the instrument 
into four identical optical channels arranged next to each other 
and supported by the same mechanical structure. 
Each optical channel is a classical focal reducer imaging spectrograph, 
with a collimator providing a parallel beam where the dispersive element 
(a grism) is inserted, and a camera that focuses the beam onto 
a 2048x4096 15 $\mu$m pixel EEV CCD. The focal plane is flattened by a field 
lens at the instrument entrance, to allow for flat multislit masks, 
and a folding mirror is inserted into the collimator, 
to fold the beam and to reduce the instrument's overall length. 
The field of view covered by each channel (generally referred to 
as a VIMOS quadrant) is approximately $7'$x$8'$, 
with a pixel scale of 0.205 arcsec/pixel.
An approximately 2 arcmin wide gap is present between quadrants.

For spectroscopic observations, six different grisms 
provide spectral resolutions ranging from $R\sim 250$ 
to 2500. Order-sorting filters are used to avoid an overlap 
between first and second grating orders. 
Laser-cut masks (one per quadrant) are used for MOS observations. 
The number of slits that can be placed on each mask varies 
from approximately 40 at high spectral resolution up to approximately 
250 at low spectral resolution. An imaging exposure acquired with VIMOS 
is required 
as the starting point of the mask design and cutting process \citep{vmmps}.

\section{VIMOS operations within the VIPERS context}
\label{vimos}
Preparing and submitting MOS observations with VIMOS requires a 
sequence of operations,
as thoroughly explained in the VIMOS User's manuals and ESO web pages. 
In service mode (which is the default observing mode) , once the pointing
location has been chosen, the information needed to carry out pre-imaging 
has to be sent to ESO, together
with the finding chart of the field.
As soon as pre-imaging data are available, the user is asked to prepare
the files needed to manufacture the masks needed for spectroscopy observations, and
send them together with the other information required (instrument configuration,exposure time,
observing sequence, etc). Mask preparation is done via
VMMPS software \citep{vmmps} distributed by ESO. Once the spectroscopic observations have 
been performed, the data can be retrieved
from ESO archive, and reduced. Finally, from the flux and 
wavelength calibrated monodimensional
spectra, redshift and other spectral quantities can be measured. \\
For normal programs, none of these operations
is particularly time consuming, nor demanding. It is when this sequence is to
be applied to a survey which foresees of the order of hundreds of pointings
and hundred thousands
spectra (as VIPERS) that the need for automatization arises. {\it Easylife} is the system we have
devised especially for VIPERS, but which can be easily adapted to other
projects requiring a high degree of data reduction automatization.
The whole reduction procedure
is based on the pipeline described in \citet{vipgi}, which we have
automatized to a very high degree as explained in section \ref{reducer}. The redshift measurement is carried out
using EZ (Easy redshift, \citealt{EZ}) in blind automatic mode. 
Even if EZ is rather efficient, especially
on this kind of data (see section \ref{EZ}), a human inspection of the
spectra is required to 
validate the measurements and possibly recover a redshift for the faintest objects.
This operation
is performed by either one or two persons (according to data quality).
Finally redshifts, together with redshift reliability flag, as well as mono
and two dimensional spectra have to be fed back to the database for
dissemination among the whole survey team. 

\section{Software requirements}
\label{requirements}
VIPERS foresees to observe about 300 VIMOS pointings in four years: each pointing observation 
is split in 5 exposures, and each exposure covers the 4 VIMOS quadrants. 
The expected data flow is thus of the order of 6000 raw data frames to be reduced and 
100000 spectra to be measured. 
For such a survey, a
semi-manual procedure as adopted for VVDS\citep{vipgi} and zCosmos \citep{zcosmos_main}  is not
 efficient enough for the data reduction: 
VVDS was made up of 99 VIMOS pointings, and zCosmos of 90 VIMOS pointings. Data reduction and redshift
measurement for both surveys
has been carried out in manual mode, using VIPGI and EZ, and had taken of the order of 5 years to be completed.
Scaling to the VIPERS case, it would translate into 15 years of efforts just to  reduce the data.
Therefore, 
the automatization of 
the processing chain, including
reduction and automatic redshift measurement, is the first requirement we had
to meet. Such an automatic pipeline must run in unsupervised mode, but has to have 
built-in quality checks on crucial steps so that the output products are
fully controlled.\\
Periodic internal data releases to the VIPERS consortium must be foreseen, to allow for
scientific exploitation even before the whole survey has been completed,
as well as periodic public releases to the whole community. This can be easily
accomplished without additional workload if, after reduction and redshift
measurement, all information is automatically entered in a database, which can be
opened, in full or in part, when data must be released.\\
Tasks like finding chart
production and mask preparation, which cannot be automatized further with
respect to the tools ESO provides, are carried out by different people,
and the
same applies to the redshift measurement task. Distributed work can be more
efficient if handy tools to get the required input (e.g. pre-imaging for preparing 
masks, reduced
monodimensional spectra to measure spectra) and send back results are used.
Web-driven upload and download procedures which take care of 
storing results and performing quality checks have to be
provided.\\
Finally, 
an adequate book-keeping must be provided for several aspects: the
managerial need of evenly distributing the workload among all partners, and
the degree of advancement of each person and each task;
information on the targets selected for the obervations has to be kept;
since the program is spread along few years, it is advisable to keep track
on when observations (both pre-imaging and spectroscopy) have been taken; 
and last, but not least, all consortium members must have the possibility to check what is the
advancement in terms of observations, data reduction, completeness, etc.
\\
Automatic reduction, database storage and book-keeping are
the basic requirements we have set for {\it Easylife}, together with 
keeping to a minimum the need for supporting man power from the `survey
reduction center`. 
On top of these requirements, it was desirable to use reduction tools already fully tuned and tested, instead of 
rewriting all the required routines from scratch. Finally, we wanted to create 
a flexible system which could be adapted to similar projects in which we are involved.
\\

\section{Software architecture}
\label{architecture}
The three main requirements described in the previous section naturally lead
to
conceive a modular sysem, where both reduction programs (usually written in C
language), databases (mySQL based) and web interfaces (developed in Java or
HTML) can live together and flawlessly interact. 
The OPTICON Future Astronomical Software Environment
(from here on, FASE, \citealt{fase1}) is a
scalable open system application framework with distributed capabilities, 
specifically studied for the astronomical software, which can by design
satisfy all these needs. 
The FASE architecture is described in \citet{fase2}, and here we recall the
fundamental concepts.
Following Figure \ref{design}, the major system elements are as follows:
\begin{itemize}
\item {\bf Presentation layer}: is the part of the system which presents 
the user the various functionalities. The user can be a human, but also
a Grid workflow, a Web browser interface, or whatever. The Presentation 
layer itself can be a Command Line Interface, a Graphical User Interface 
or a Web interface.
\item {\bf Application layer}: the application layer is used to implement top level applications.
The application layer can be anything which can drive the execution
framework to execute components, for example, Python, Java, a GUI, or a work-flow engine of some sort.
\item {\bf Execution framework}: provides the functionality needed to execute components,
including capabilities such as component registration and management,
distributed execution, scalability, messaging, logging, and so forth. Different execution frameworks, each
one having different capabilities, can be implemented.
\item {\bf Container}: components execute within a container which defines the life cycle
and runtime environment seen by the component. The container is the interface
between the execution framework and an individual component.
\item {\bf Components}: a component is a computational object, with one or more service
methods, which can be plugged into the framework. Components are grouped
into component packages and provide most of
the functionality of the system.
\end{itemize}

The component-framework architecture outlined here is a modular
architecture in which  the major elements of
the system can be used separately, as stand alone packages, or can be
integrated into other frameworks. 
The advantage
of a modular architecture is that the major elements of the system can evolve independently,
making it easier to use new technology as it becomes available.
Developing {\it Easylife}, we have made full use of the modularity FASE provides: 
some elements (the \textit{Reducer} and to a certain extent the \textit{Unpacker})
were pre-existing and we have just plugged them in the global system after
having built the appropriate container. At the same time we have
been able to extend the system capabilities 
by adding extra components for the data reduction of other
instruments (like LUCI and MODS at the LBT, see section \ref{LBT}).

\section{{\it Easylife} building blocks}
\label{blocks}
Following the architectural concept of FASE, the different tasks deriving
from the requirements outlined in section \ref{requirements}
are handled through a dedicated {\it Easylife} 
component and/or GUI. 
The survey status can be monitored and managed through an administrative
web site, which is also used to provide the public Web pages of the VIPERS
survey. Data ingestion, organization and reduction tasks, together
with automatic redshift measurement, are carried out by dedicated
tools running on a beowulf cluster. Finally, the results database is 
based on mySQL and accessed through a dedicated GUI.
All these parts communicate with an
administrative SQL-based database, which keeps track of the global status of
the survey, and all together constitute the
{\it Easylife} system.

\subsection{VIPERS administrative web site}
\label{admin}
The {\it administrative web site} is the uppermost presentation layer of {\it Easylife}. It 
allows one to monitor the survey status, access all survey-related side products,
as outlined below, and retrieve data.

The VIPERS administrative web site is built on top of a Web application
framework running on a Jakarta Apache server. It allows one to serve normal static HTML
pages as well as dynamic pages. VIPERS pages are built upon a template system integrated within
the Web application framework, which ensures homogeneity of the layout. 
The Web application framework is fully integrated within the
{\it Easylife} management system, and directly accesses the underlying SQL database,
which contains all the relevant information for the survey monitoring. 
It is structured to have different access levels: a public part (http://vipers.inaf.it), which
describes the survey goals, shows the team composition and will contain a summary
of the most relevant results; a team restricted part and an administrative
part, with access restricted to the PI and the administration team.
Through the private part, each member 
of the team can retrieve the information
he/she may need, e.g. 

\begin{figure*}
  \includegraphics[clip=true,scale=0.7]{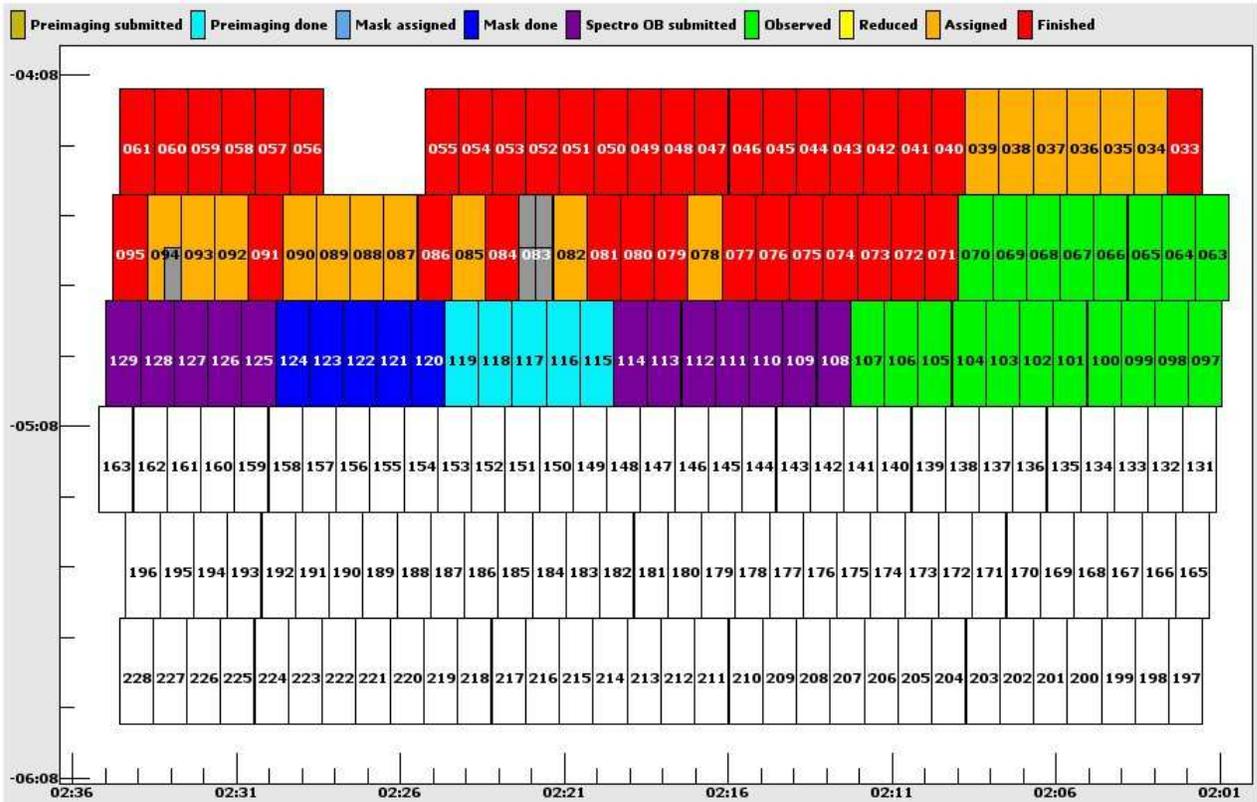}
  \caption{Example of panel showing the status of the observations. The
           graph displays the pointings placing them in the correct
           coordinates and coloring each pointing with a different
           color depending on its current status. The status ranges from 
           pre-imaging submitting up to data validation assignment,
           with a final status assigned when the processing of the pointing 
           has been definitely closed.}
  \label{pntgs}
\end{figure*}

\begin{figure}
  \plotone{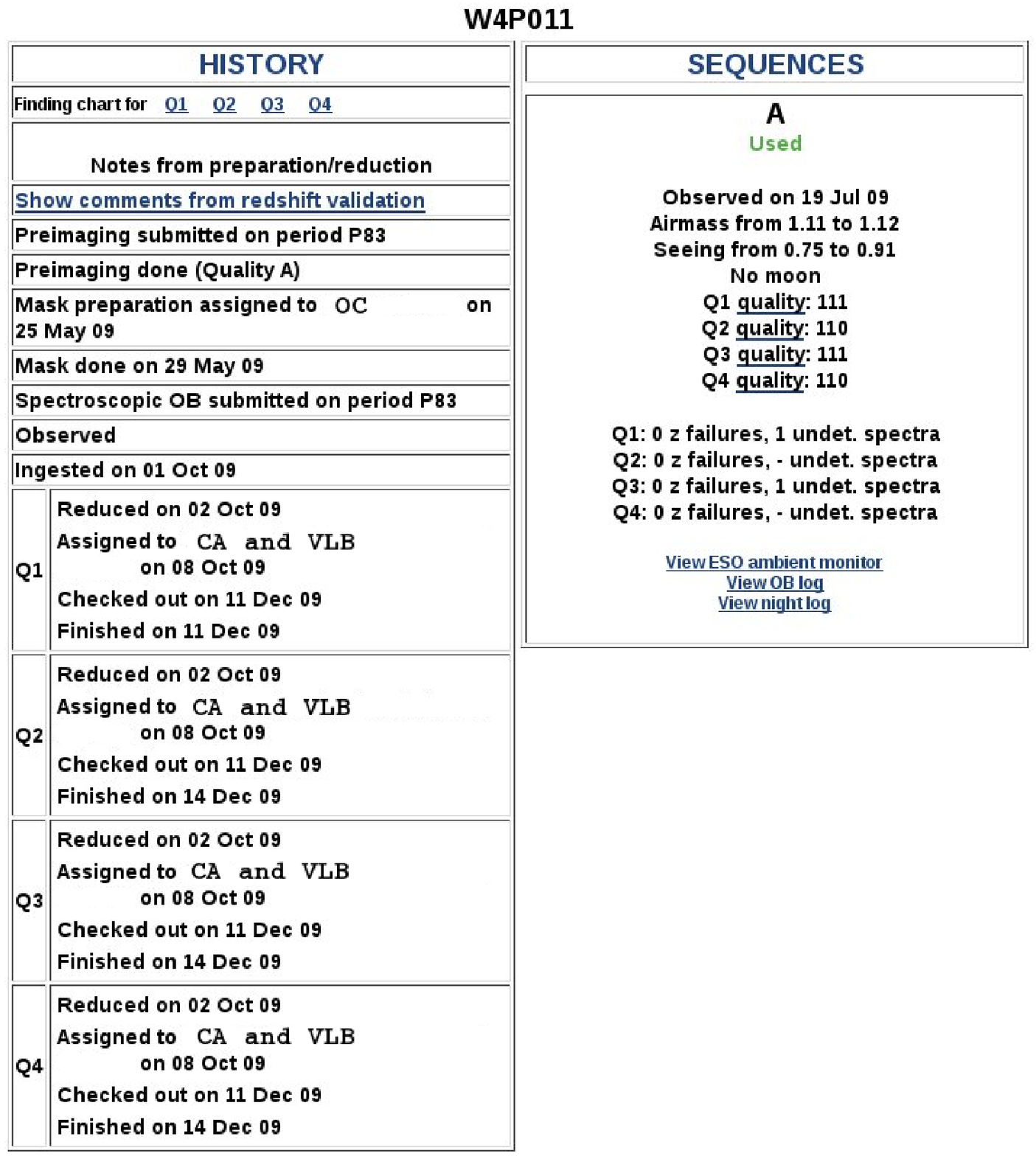}
  \caption{Example of panel showing the observation and reduction details for one pointing}
  \label{pnt_detail}
\end{figure}

\begin{itemize}
 \item inspect how the survey is advancing. An example is given in 
Figure \ref{pntgs} for the CFHTLS-W1 area. The different colors indicate the
different advancement status of each pointing (green for observed, yellow
for reduced, red for fully measured,  etc.). For each pointing, relevant
information such as date of observation, metereological conditions during
observations (through a link to observation logs provided by ESO),
data quality, are accessible by clicking on the pointing
itself (see Figure \ref{pnt_detail}). These figures are created {\it on the
  fly} from the SQL database holding all survey information, and thus are
always automatically up to date.

\item Connect to the database system, providing the photometric parent
       catalogs and the catalogs with the scientific information
       extracted from the spectra. The database system is based on DART
       software \citep{dart}, a Web interface which allows one to
       query catalogs and access their associated data products (see
       section \ref{dart}).
\item Have access to project documentation and meetings minutes,
as well as to the
VIPERS science wiki pages related to different internal projects or
working groups. 
\item upload any VIPERS related publication, and look at publications or
  presentations given by team members
\item retrieve the data for mask preparation or redshift
  measurement and upload the results  
\end{itemize}

The administrative pages are reserved to the PI or project admistrator to 
\begin{itemize}
 \item Assign the VIMOS mask preparation to the team members 
 \item Once data have been reduced,
       assign the redshift measurement validation to the different team
       members. 
 \item Make new data releases, freezing the current status of the
       spectroscopic catalogs and labeling them with a custom tag.
       Some statistics are then produced summarizing the survey status
       and outcome.
\item keep track of the ``service'' work done by each team member, to avoid
  overload of some with respect to others.
\end{itemize}

\subsection{Data ingestion and reduction}
\label{datareduction}
While the administrative web site allows one to handle the global phases of
the survey process, data management and reduction are performed by a restricted {\it data reduction
group} through
a dedicated Graphical User Interface.
Such GUI handles three main software elements, each of which is
dedicated to a specific set of reduction and management tasks:

\begin{itemize}
  \item \textit{Unpacker} (section \ref{unpacker}): unpacks the raw data and prepares them for 
        ingestion in the reduction system;
  \item \textit{Organizer} (section \ref{organizer}): fills the database containing the pointing
        information and organizes the data in a pre-defined structure,
        classifying each file by its attributes;
  \item \textit{Reducer} (section \ref{reducer}): reduces the raw data in order to produce
        mono-dimensional wavelength and flux calibrated spectra for
        each target object and measures the spectroscopic redshifts.
\end{itemize}

The \textit{Unpacker}, \textit{Organizer} and \textit{Reducer} are used
through the GUI in a seamless way, proposing the user to choose: (a)
the project to be handled; (b) the raw data to be unpacked; (c) the
pointings to be reduced and the related files to be used for the
reduction; (d) launch the reduction process.

\subsubsection{Data preparation}
\label{unpacker}

The \textit{Unpacker} is the {\it Easylife} software element dedicated to ingest raw
data into the reduction system. 
EasyLife has been conceived with the aim of being usable for several projects
and several spectrographs. The purpose of the unpacker is to analyse the
raw data it receives, discard whatever is not needed/wanted 
and add to the header of the raw data files some conventional keywords,
which will allow the data to be classified according to the project/instrument they
belong to. 
The exact behaviour of the Unpacker is driven by configuration settings, which
essentially indicate where, and in which form, the information required is
contained in the raw data. 
At the end of the process, each raw data file contains standard hierarchical FITS
keywords  which contain the main information required for
classifying the file independently of the instrument: for example, the disperser
used, the target name, the instrument name, the airmass, and others.
The file is also renamed following a "human readable" syntax which allows one to
immediately identify whether it is a scientific exposure, a flat field, which
is the target and with which disperser it has been observed.

{\it Easylife} hierarchical FITS keywords provide a conventional set of information
irrespective of the instrument which has produced the data. This information is what
the \textit{Organizer} needs to classify the data.

It is worth noting that this approach to data ingestion allows one to use {\it Easylife}
for different projects and even instruments: for each target aplication (be it
a survey with VIMOS, or several observations with another spectrograph), it is
sufficient to configure a different \textit{Unpacker} to customize {\it Easylife} 
for projects
very different from the VIPERS survey it has been devised for. In section
\ref{LBT} we will show how we have already used
{\it Easylife} for other projects.

\subsubsection{Data organization}
\label{organizer}
Once the data contain a set of standard information in a standard format, 
they can be easily classified and
organized according to their content. The classification is stored in a mySQL
database (which is also accessed by the web interface, see section \ref{admin}), while 
the data management operations are performed through the
\textit{Organizer}, which 
provides the data organization and administration functions. 
The \textit{Organizer} handles multiple projects (VIPERS
application is one project), providing a separate work space for each
one.
A project work space consists of: 1) a data storage area, which points
to a well defined directory structure; 2) a set of database tables: 
the table holding the files attributes and their reduction status,  the table
collecting the administrative information concerning the pointings 
(or targets) and their global status, and the table containing the information on the
calibration files and their validity range.
Every inquiring operation on the files and/or on the survey management
process is performed by the different {\it Easylife} components accessing the
\textit{Organizer}. The \textit{Organizer} is thus the main element that
allows one to orchestrate the entire management system. 

\subsubsection{Data reduction and quality control}
\label{reducer}
\begin{figure}
  \plotone{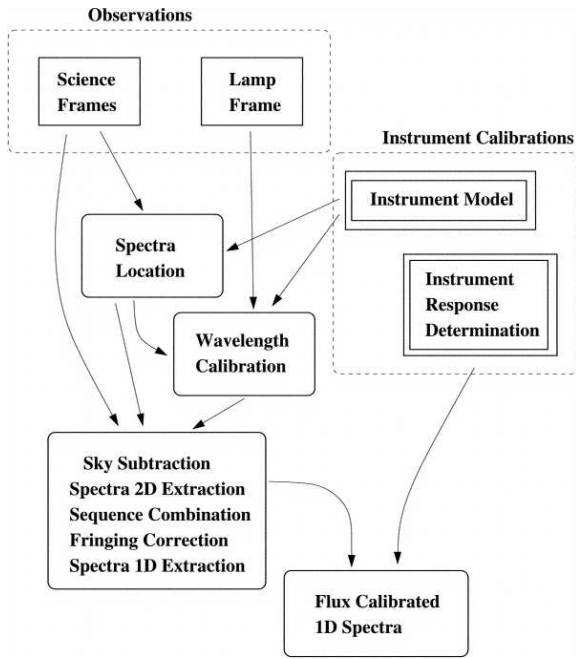}
  \caption{Block diagram summarizing the main steps involved 
in the reduction of VIMOS data \citep{vipgi}}
  \label{vipgischeme}
\end{figure}

The data reduction is performed with a special {\it Easylife}
software component (the \textit{Reducer}), which provides an
automatic pipeline system equipped with a specific plug-in for the VIMOS
instrument. Thanks to FASE
distributed execution engine, the \textit{Reducer} is able to
process multiple observations at the same time, submitting the
reduction processes to a Beowulf cluster.
The reduction steps and underlying recipes are described in \citet{vipgi}, and
recalled in Figure \ref{vipgischeme}.
Briefly, the implemented global data reduction scheme is a 
fairly
traditional one, broadly following the one implemented by the IRAF 
long‐slit package: 1) location of spectral traces on the raw frames, 2) computation of the Inverse 
Dispersion Solution for each spectral trace, 3) sky subtraction
on the non-calibrated data, 4) two dimensional extraction of spectra and
application of the wavelength calibration, 5) combination of sequence of observations
6) extraction of mono-dimensional spectra and correction for the isntrument
sensitivity function (flux calibration).   
A special effort was made to achieve a 
very high efficiency during the repeated application of this scheme 
to the large set of VVDS data by tailoring all aspects of the 
data reduction scheme to the specific characteristics of VIMOS. Still, 
the various reduction functions are general enough that they can 
be adapted for the reduction of data produced by any MOS spectrograph, 
with a minimal effort (see section \ref{LBT} and \citealt{fvipgi}).
Such recipes, in their original form, formally always end successfully, 
but this does not 
automatically mean that the result meets the degree of accuracy required by 
the specific scientific need. For example, a spectrum can be successfully wavelength calibrated,
but the wavelength calibration accuracy is of the order of 1 pixel. This is clearly not enough if
the redshift accuracy required is much higher than that.
In the past, reduction results were always manually checked and, when required, data were 
reduced again in order to improve results.
Given the high data flow of VIPERS (6000 raw frames), 
the fully automated pipeline must also assure that the reduction results are 
scientifically exploitable.
For these reasons, on top of the reduction flow described in \citet{vipgi}, we 
added some quality check steps. 
Every time one of such quality checks is not
satisfied, the reduction process is stopped and human intervention is required.
We have explored the parameter space of each step in order to find
the minumum (maximum) value above (below) which VIPERS data are scientifically 
usable. Such limits are stored in a configuration file.
The quality checks we perform, together with the
the adopted limits are the following:

\begin{enumerate}

\item \textit{Check on spectra location}. Each VIPERS observation consists
of several exposures, possibly spread over different nights. It is well
known that VIMOS suffers from a flexure problem (only recently fixed thanks to
an Active Flexure Compensator, see \citealt{flexure_comp}) so that the
location of the dispersed spectra on the different
exposures can differ by few pixels from the expected positions. For this
reason, the task computes the expected spectrum border position and
compares it with the real detected spectrum border. The median of this displacement
is requested not to exceed 1.5 pixels for 5\% of the spectra in one VIMOS quadrant. 
If these conditions are not satisfied, the expected position is
not accurate enough to guarantee a good spectrum tracing and therefore exctraction in all exposures
of the same field, 
and the procedure is stopped to allow for a manual adjustment
of the slit position first guess.
\item \textit{Check of wavelength calibration}. Using the Inverse 
Dispersion Solution derived by the pipeline,
the expected position of each reference spectrum line in each slit is computed. Such expected
position is then compared to the actual arc line position as measured from the
raw data and the difference between expected and observed position is
computed. For each slit, the RMS of such differences is also computed.
The quality control is successfull when all the following conditions are satisfied:
        \begin{itemize}
         \item the median of the RMS distribution using all slits is
               not larger than 0.2 pixels;
         \item for each slit, the RMS is not higher than 0.1 pixels and lower 
           than 0.3 pixels. This condition must be satisfied at least by 90\% of the
           slits.
         \item in each slit, the minimum number of arc lines used to fit the
           Inverse Dispersion Solution is at least 9.
         \item the bluest and reddest visible arc lines are within 2.5$\sigma$
           from the best fit for at leat 90\% of the slits
        \end{itemize}

\item \textit{Detected targets}. Once data have been reduced and
  monodimensional spectra
  extracted, the number of detected targets is computed. In general, given the exposure time and the limiting magnitude of
  the survey, we expect
  a detection rate above 90\%. If such threshold is not reached, it is usually the 
  signal that the metal mask, on which the slits are carved, was badly positioned
  on the focal plane (an event which may occur, see also \citealt{flexure_comp})
  or of bad observing conditions. In this last case, also
  the observation quality flag (see below) independently indicates bad quality data.

\item \textit{Quality flag}. When all exposures belonging to the same pointing
have been reduced and
  combined together, a check on some environmental parameters which can affect the
        quality of the data is performed (see Garilli et al.
        2008 for details). In particular, we check the mean PSF as measured
        from the reduced image, the measured sky brightness, and the object
        centering in the slit. These  three quality
        parameters can score 1 (good) or 0 (bad) and they are combined together
        in order to produce a final reduction quality flag

\end{enumerate}

\subsection{Redshift measurement}
\label{EZ}
Once the data are fully reduced, they are ingested into a blind redshift measurement 
pipeline provided by
EZ, fully described in \citet{EZ}. EZ has been developed within the
VVDS project to help in redshift measurement from optical spectra.
The basic idea is to allow the user to
combine the available functions in 
the most appropriate way for the data at hand, thus building new user
defined functions and methods. At the upmost level, a redshift measurement
{\it decision tree} can be built, which mimicks the decision
path followed by an astronomer to get to the measure of the redshift. 
Complete automation of the redshift measurement process can be tricky
when spectra are noisy (as they always are at the faint limit of a
survey)
or in presence of artifacts such as fringing
correction residuals, so that  it
is by no means guaranteed, a priori, that the best solution proposed
is also a correct solution. 
For this reason, EZ also computes
a reliability flag which
summarizes the goodness of the solution proposed. As for the redshift, also
the reliability flag computation is performed
mimicking the  kind of logical reasoning applied by an
astronomer when trying to
evaluate if a redshift is reliable or not. 
The implemented flagging system is rather conservative, as
demonstrated in \citet{EZ}.
EZ can be used both interactively, or totally blindly in unsupervised mode, which is the mode
we have adopted within {\it Easylife}. 

The redshifts thus obtained are compared for consistency with
the photometric redshifts, and an approprate decimal flag is added to the 
reliability flag provided by EZ.
This particular final step of the reduction is applicable in the case of VIPERS, but 
could be not applicable in other cases. The modularity of {\it Easylife}
allows to switch on or off any reduction step, according to the needs.
The final redshifts approval is
formalized after a human check. The reduced data are submitted
to the survey team members who are in charge of the redshift validation, who have 
at their disposal: the
mono and two-dimensional object spectra, together with their associated sky and
noise spectra, the output of the automatic measurement, with associated flag,
and the information whether such measurement agrees or not with the
photometric redshift (within the photometric redshift error). 
In section \ref{performances} we will show how the automatic
redshift measurements performs on the VIPERS data.

\subsection{Survey Database}
\label{dart}
Once redshifts have been humanly validated and uploaded to the survey web site, they automatically enter
the spectroscopic data\-base, together with the other scientifically interesting
quantities such as the object magnitude in the selection band and its coordinates.
The database also hosts the parent photometric catalog, containing {\it ugriz}
magnitudes from the CFHTLS survey and photometric redshifts.
Periodically (tipically on a yearly basis) the spectroscopic catalog is frozen in a
release, which is made available to the whole team for scientific analysis and checks.

{\it Easylife} allows one to access the photometric parent
catalogs and the spectroscopic catalogs through an embedded DART Web
interface installation. As described in \citet{dart}, DART gives a
per-user access to the data allowing to query catalogs, filter data by
placing conditions on the column values (even complex expressions), view the
results and export them to private user files stored in the remote
data server. DART also allows to make simple plots or retrieve the
data products related to the catalogs, as the mono-dimensional spectra
resulting from the reduction process or any other ancillary data
product (image thumbnails of different bands, links to external web sites,
documents, etc.). The software supports access to more than one catalog
at a time (e.g. for multi-band usage):

\begin{itemize}
  \item in parallel, namely querying each catalog singularly at the
        same time;
  \item as a couple linked by a pre-built correlation table released by
        the data managers;
  \item as a single virtual table, which allows to view the result of 
the pure correlation by objects ID among
        several catalogs ;
\end{itemize}

DART supports also IVOA SSA protocol for the spectra access, IVOA
SIA protocol for images access and ConeSearch protocol for
catalogs access  (http://www.ivoa.net/Documents/), allowing to open a gate 
towards the Virtual
Observatory facilities for VIPERS data. 
DART allows to give different access privileges to different user classes, 
so that at  the same time one can have a public part,
a team reserved part, containing the most recent release, and a restricted
part not yet released to the team, containing the data being accumulated in 
after the last team release.

\section{{\it Easylife} performance}
\label{performances}
The {\it Easylife} reduction blocks detailed above, coupled with the automatic redshift measurement, 
are very efficient: the full chain, including all the automatic quality checks, 
requires about 40 minutes of computation time per pointing (each VIPERS
pointing containing of the order of 320 spectra), without supervision 
or human intervention.

\subsection{Data reduction performance}
Human intervention is required when one of the quality checks described in section \ref{reducer}
fails and the procedure is stopped. In Table \ref{reduction_stats}, we give the failure rate
of the automatic reduction procedure 
we have experienced in the first 4x113=452 quadrants of the VIPERS survey.  
For  92\% of the observations, the automatic reduction ran smoothly
without human intervention and the data satisfied all quality checks. 
In only 2.5\% of the data (i.e. 11 quadrants) the automatic procedure has
failed either to automatically locate spectra (9 quadrants) or to derive a
good wavelength calibration solution (within the limits set in the quality control configuration file).
\\
The check which fails the most (5.5\% of the times, i.e. 24 quadrants) 
is the one on the number of detected sources. Cross correlation of the quality parameter with
these quadrants shows that in 10 over 24 cases observing conditions below average
are responsible for the lower than average detection rate, while other observational hardware 
problems (e.g. guide lost during observation, field partially obscured by the
guide probe, bad maskinsertion) account for the low detection rate of 10 other quadrants. 
In only 4 cases (less than 1\%) the low detection rate seems to be due to local problems
in the photometric catalogue, affected by the presence of  a bright star, or
by a poor astrometric solution when preparing masks, which may affect the corners of
the field. Overall, our quality control proves to be reliable and allows us to quickly
spot data below average quality.
This information is not only useful {\it per se} but also to assign pointings for redshift measurement
check: while higher quality data can be checked by one person only, the lower
quality ones are systematically looked at by two different people.
\\
\subsection{Automatic redshift measurement performance}
All VIPERS redshifts have been manually validated, as it had been done
for the VVDS and the zCosmos surveys (\citet{vvds_main},\citet{zcosmos_main}). 
In \citet{EZ}, it has been showed that EZ, used in blind mode, had a
measurement success rate of 95\% on simulated data, while on the VVDS and
zCosmos surveys the success rate was $\sim 70\%$ on the whole sample, rising
to $90\%$ for redshifts classified as very secure by astronomers.
In Table \ref{perfTable} we summarize the results obtained on the first $\sim 36000$ detected targets
belonging to the 113 VIPERS pointings mentioned above.
The redshift flag scheme implemented in EZ mimicks the one adopted for 
the VVDS and the zCosmos surveys, i.e.
\begin{itemize} 
\item
    flag 4: a 100\% secure redshift, with high SNR spectrum and
    obvious spectral features supporting the redshift measurement;  
\item
    flag 3: a 90\% secure redshift, strong spectral features; 
\item
    flag 2: a 75\% secure redshift measurement, several features in support
    of the measurement; 
\item
    flag 9: only one secure single spectral feature in emission,
    typically interpreted as [OII]3727 Å, or $\rm H\alpha$.
\item
    flag 1: a 50\% reliable redshift measurement, based on weak spectral features
    and continuum shape;  
\item
    flag 0: no reliable redshift measurement possible; 
\end{itemize}
In the table, results are subdivided by automatic reliability flag.
For
each redshift automatically measured by EZ, and for each automatic flag (column 1), 
the table shows the number of spectra for which EZ has measured a redshift assigning that particular 
reliability flag (column 2), the number of spectra for which redshift has been confirmed 
by the astronomers (column 3), 
 and the resulting success rate (column 4). 
The results shown in Table \ref{perfTable} are in line with those
already obtained for the VVDS Wide survey: overall, the automatic measurement has
been confirmed for  76\% of the spectra, confirmation rising to 94\%
for automatic flags 3 and 4.  
\\
Table \ref{perfTable} also shows that the automatic flag is more restrictive than the
human one, as already stated in \citet{EZ}: 52\% of the 
redshifts flagged as 0 (unreliable) by EZ have been confirmed by astronomers.
Table \ref{real_flag_table} compares the automatically assigned flags with
the human ones, when the automatic redshift has been confirmed.
We can see that flags 3 and 4 have been confirmed 81\% of the times, 
flags 2-9 55\% and flags 1 19\% of the times, while automatic flags 0 became flags 3-4 in 27\%
of the cases, again supporting the conservativeness of the automatic flag assigment.

\section{Using Easylife for LBT data}
\label{LBT}
The modular approach of {\it Easylife} has allowed  us to easily adapt it to other, totally different
projects. Currently, it is used within the framework of the Italian LBT (Large Binocular Telescope, \citealt{lbt}) Data Center to reduce
all spectroscopic observations obtained with either MODS (Multi-Object Double Spectrographs, \citealt{mods}) or LUCI
(LBT NIR spectroscopic Utility with Camera and Integral-field unit, \citealt{lucifer}) during the Italian observing time.
Being MODS a multiobject slit based spectrograph operating in the visible range, similar to VIMOS in its concept, 
adaptation of the reduction part has been straightforward, the required intervention being limited to the
development of the instrument dedicated part of the {\it Unpacker}.
LUCI is a multiobject spetrograph working in the NIR J,H and K bands. Therefore, on top
of a dedicated  {\it Unpacker}, some more work on the reduction recipes has been
performed, to comply with the specific peculiarities of the NIR spectroscopic data (e.g. the 
much more delicate problem of the sky subtraction). But the main difference between the reduction center for
a large scale
survey, like VIPERS, and the reduction center for a whole community, like
the LBT Italian Data Center, resides in the different services the two centers must provide. While in the frst
case, data are acquired with the same instrument configuration, which makes reduction easier, but
a number of other tasks are required (logging, data base, etc), in the second case data are acquired 
with a variety of
instrument configurations, satisfying a variety of  scientific needs, and the reduction chain
must be able to cope with such diversities. On the other hand, the management part, as well as 
data products distribution, is minimal: the only two actions required are to keep
track of which data have been reduced and what remains to be done, on one side, and make available the reduced data
to the PIs on the other side. 
In spite of these fundamental differences, {\it Easylife} can handle both cases: in the LBT application, the WEB part 
has been suppressed, and the management data base is structured in a different way. The reduction chain
is more versatile with several branches according to instrument mode, while the redshift measurement part
is suppressed. Adaptation of {\it Easylife} from the VIPERS survey case to the LBT service data center case
has required only few months work (mostly devoted to the implementation of the NIR dedicated
reduction recipes), thanks
to the modular approach followed since the beginning, as well as to the carefull design of the
basic architecture. 

\section{Summary}
\label{summary}
{\it Easylife} is the automatic data reduction and management system set up for the
VIPERS survey. {\it Easylife} allows to automatically reduce large amount of data
in a timely way and performs reliable quality controls on the data quality (namely the
observing conditions) and on data reduction. 
The reduction chain ends with automated redshift measurements\\
The automatic quality controls inserted in the pipeline have shown
that reduction is successfull in $>95\%$ of the cases, when observing conditions
are within specifications. The observations not satisfying the requested observing constraints
are automatically spotted and account for the vast majority of automatic reduction failures.
{\it Easylife} also comprises project support tools, a survey advancement logging system, and
data access through a dedicated data base. 
The underlying FASE software environ\-ment adop\-ted allows a smooth interaction between the database,
the core of the reduction system and the publicly exposed web interface, as well as distributed 
computing on a beowulf cluster.\\
Presently, {\it Easylife} is also used in the framework of the LBT spectroscopic data reduction center, providing
PIs with fully reduced and calibrated spectra, see e.g. \citet{luciReduction}.

\begin{acknowledgements}
We thank D.Maccagni for having named ``{\it Easylife}'' the automated data reduction system,and
the whole VIPERS team for feedback and human verification that allowed checking the
performance. We also thank an anonymous referee for carefully reading the manuscript.
This work has been
partially suppported by PRIN INAF 2009, PRIN INAF 2011, OPTICON wg 9.2,   
INAF Project Department via the Information Systems unit.
\end{acknowledgements}
\bibliographystyle{aa}
\bibliography{ms}

\begin{thebibliography}{41}
\expandafter\ifx\csname natexlab\endcsname\relax\def\natexlab#1{#1}\fi

\bibitem[{{Abazajian} {et~al.}(2009){Abazajian}, {Adelman-McCarthy},
  {Ag{\"u}eros}, {Allam}, {Allende Prieto}, {An}, {Anderson}, {Anderson},
  {Annis}, {Bahcall}, {Bailer-Jones}, {Barentine}, {Bassett}, {Becker},
  {Beers}, {Bell}, {Belokurov}, {Berlind}, {Berman}, {Bernardi}, {Bickerton},
  {Bizyaev}, {Blakeslee}, {Blanton}, {Bochanski}, {Boroski}, {Brewington},
  {Brinchmann}, {Brinkmann}, {Brunner}, {Budav{\'a}ri}, {Carey}, {Carliles},
  {Carr}, {Castander}, {Cinabro}, {Connolly}, {Csabai}, {Cunha}, {Czarapata},
  {Davenport}, {de Haas}, {Dilday}, {Doi}, {Eisenstein}, {Evans}, {Evans},
  {Fan}, {Friedman}, {Frieman}, {Fukugita}, {G{\"a}nsicke}, {Gates},
  {Gillespie}, {Gilmore}, {Gonzalez}, {Gonzalez}, {Grebel}, {Gunn},
  {Gy{\"o}ry}, {Hall}, {Harding}, {Harris}, {Harvanek}, {Hawley}, {Hayes},
  {Heckman}, {Hendry}, {Hennessy}, {Hindsley}, {Hoblitt}, {Hogan}, {Hogg},
  {Holtzman}, {Hyde}, {Ichikawa}, {Ichikawa}, {Im}, {Ivezi{\'c}}, {Jester},
  {Jiang}, {Johnson}, {Jorgensen}, {Juri{\'c}}, {Kent}, {Kessler}, {Kleinman},
  {Knapp}, {Konishi}, {Kron}, {Krzesinski}, {Kuropatkin}, {Lampeitl},
  {Lebedeva}, {Lee}, {Lee}, {Leger}, {L{\'e}pine}, {Li}, {Lima}, {Lin}, {Long},
  {Loomis}, {Loveday}, {Lupton}, {Magnier}, {Malanushenko}, {Malanushenko},
  {Mandelbaum}, {Margon}, {Marriner}, {Mart{\'{\i}}nez-Delgado}, {Matsubara},
  {McGehee}, {McKay}, {Meiksin}, {Morrison}, {Mullally}, {Munn}, {Murphy},
  {Nash}, {Nebot}, {Neilsen}, {Newberg}, {Newman}, {Nichol}, {Nicinski},
  {Nieto-Santisteban}, {Nitta}, {Okamura}, {Oravetz}, {Ostriker}, {Owen},
  {Padmanabhan}, {Pan}, {Park}, {Pauls}, {Peoples}, {Percival}, {Pier}, {Pope},
  {Pourbaix}, {Price}, {Purger}, {Quinn}, {Raddick}, {Fiorentin}, {Richards},
  {Richmond}, {Riess}, {Rix}, {Rockosi}, {Sako}, {Schlegel}, {Schneider},
  {Scholz}, {Schreiber}, {Schwope}, {Seljak}, {Sesar}, {Sheldon}, {Shimasaku},
  {Sibley}, {Simmons}, {Sivarani}, {Smith}, {Smith}, {Smol{\v c}i{\'c}},
  {Snedden}, {Stebbins}, {Steinmetz}, {Stoughton}, {Strauss}, {Subba Rao},
  {Suto}, {Szalay}, {Szapudi}, {Szkody}, {Tanaka}, {Tegmark}, {Teodoro},
  {Thakar}, {Tremonti}, {Tucker}, {Uomoto}, {Vanden Berk}, {Vandenberg},
  {Vidrih}, {Vogeley}, {Voges}, {Vogt}, {Wadadekar}, {Watters}, {Weinberg},
  {West}, {White}, {Wilhite}, {Wonders}, {Yanny}, {Yocum}, {York}, {Zehavi},
  {Zibetti}, \& {Zucker}}]{sdss7}
{Abazajian}, K.~N., {Adelman-McCarthy}, J.~K., {Ag{\"u}eros}, M.~A., {et~al.}
  2009, \apjs, 182, 543

\bibitem[{{Auld} {et~al.}(2006){Auld}, {Minchin}, {Davies}, {Catinella}, {van
  Driel}, {Henning}, {Linder}, {Momjian}, {Muller}, {O'Neil}, {Sabatini},
  {Schneider}, {Bothun}, {Cortese}, {Disney}, {Hoffman}, {Putman}, {Rosenberg},
  {Baes}, {de Blok}, {Boselli}, {Brinks}, {Brosch}, {Irwin}, {Karachentsev},
  {Kilborn}, {Koribalski}, \& {Spekkens}}]{ages}
{Auld}, R., {Minchin}, R.~F., {Davies}, J.~I., {et~al.} 2006, \mnras, 371, 1617

\bibitem[{{Bertin} {et~al.}(2002){Bertin}, {Mellier}, {Radovich}, {Missonnier},
  {Didelon}, \& {Morin}}]{terapix}
{Bertin}, E., {Mellier}, Y., {Radovich}, M., {et~al.} 2002, in Astronomical
  Society of the Pacific Conference Series, Vol. 281, Astronomical Data
  Analysis Software and Systems XI, ed. {D.~A.~Bohlender, D.~Durand, \&
  T.~H.~Handley}, 228

\bibitem[{{Bottini} {et~al.}(2005){Bottini}, {Garilli}, {Maccagni}, {Tresse},
  {Le Brun}, {Le F{\`e}vre}, {Picat}, {Scaramella}, {Scodeggio}, {Vettolani},
  {Zanichelli}, {Adami}, {Arnaboldi}, {Arnouts}, {Bardelli}, {Bolzonella},
  {Cappi}, {Charlot}, {Ciliegi}, {Contini}, {Foucaud}, {Franzetti}, {Guzzo},
  {Ilbert}, {Iovino}, {McCracken}, {Marano}, {Marinoni}, {Mathez}, {Mazure},
  {Meneux}, {Merighi}, {Paltani}, {Pollo}, {Pozzetti}, {Radovich}, {Zamorani},
  \& {Zucca}}]{vmmps}
{Bottini}, D., {Garilli}, B., {Maccagni}, D., {et~al.} 2005, \pasp, 117, 996

\bibitem[{{Coil} {et~al.}(2004){Coil}, {Davis}, {Madgwick}, {Newman},
  {Conselice}, {Cooper}, {Ellis}, {Faber}, {Finkbeiner}, {Guhathakurta},
  {Kaiser}, {Koo}, {Phillips}, {Steidel}, {Weiner}, {Willmer}, \&
  {Yan}}]{deep2}
{Coil}, A.~L., {Davis}, M., {Madgwick}, D.~S., {et~al.} 2004, \apj, 609, 525

\bibitem[{{Colless} {et~al.}(2001){Colless}, {Dalton}, {Maddox}, {Sutherland},
  {Norberg}, {Cole}, {Bland-Hawthorn}, {Bridges}, {Cannon}, {Collins}, {Couch},
  {Cross}, {Deeley}, {De Propris}, {Driver}, {Efstathiou}, {Ellis}, {Frenk},
  {Glazebrook}, {Jackson}, {Lahav}, {Lewis}, {Lumsden}, {Madgwick}, {Peacock},
  {Peterson}, {Price}, {Seaborne}, \& {Taylor}}]{2df}
{Colless}, M., {Dalton}, G., {Maddox}, S., {et~al.} 2001, \mnras, 328, 1039

\bibitem[{{Davis} {et~al.}(1982){Davis}, {Huchra}, {Latham}, \&
  {Tonry}}]{Davis1980CfA1}
{Davis}, M., {Huchra}, J., {Latham}, D.~W., \& {Tonry}, J. 1982, \apj, 253, 423

\bibitem[{{Drinkwater} {et~al.}(2010){Drinkwater}, {Jurek}, {Blake}, {Woods},
  {Pimbblet}, {Glazebrook}, {Sharp}, {Pracy}, {Brough}, {Colless}, {Couch},
  {Croom}, {Davis}, {Forbes}, {Forster}, {Gilbank}, {Gladders}, {Jelliffe},
  {Jones}, {Li}, {Madore}, {Martin}, {Poole}, {Small}, {Wisnioski}, {Wyder}, \&
  {Yee}}]{WiggleZ}
{Drinkwater}, M.~J., {Jurek}, R.~J., {Blake}, C., {et~al.} 2010, \mnras, 401,
  1429

\bibitem[{{Eisenstein} {et~al.}(2001){Eisenstein}, {Annis}, {Gunn}, {Szalay},
  {Connolly}, {Nichol}, {Bahcall}, {Bernardi}, {Burles}, {Castander},
  {Fukugita}, {Hogg}, {Ivezi{\'c}}, {Knapp}, {Lupton}, {Narayanan}, {Postman},
  {Reichart}, {Richmond}, {Schneider}, {Schlegel}, {Strauss}, {SubbaRao},
  {Tucker}, {Vanden Berk}, {Vogeley}, {Weinberg}, \& {Yanny}}]{sdss1}
{Eisenstein}, D.~J., {Annis}, J., {Gunn}, J.~E., {et~al.} 2001, \aj, 122, 2267

\bibitem[{{Garilli} {et~al.}(2010){Garilli}, {Fumana}, {Franzetti}, {Paioro},
  {Scodeggio}, {Le F{\`e}vre}, {Paltani}, \& {Scaramella}}]{EZ}
{Garilli}, B., {Fumana}, M., {Franzetti}, P., {et~al.} 2010, \pasp, 122, 827

\bibitem[{{Garilli} {et~al.}(2008){Garilli}, {Le F{\`e}vre}, {Guzzo},
  {Maccagni}, {Le Brun}, {de la Torre}, {Meneux}, {Tresse}, {Franzetti},
  {Zamorani}, {Zanichelli}, {Gregorini}, {Vergani}, {Bottini}, {Scaramella},
  {Scodeggio}, {Vettolani}, {Adami}, {Arnouts}, {Bardelli}, {Bolzonella},
  {Cappi}, {Charlot}, {Ciliegi}, {Contini}, {Foucaud}, {Gavignaud}, {Ilbert},
  {Iovino}, {Lamareille}, {McCracken}, {Marano}, {Marinoni}, {Mazure},
  {Merighi}, {Paltani}, {Pell{\`o}}, {Pollo}, {Pozzetti}, {Radovich}, {Zucca},
  {Blaizot}, {Bongiorno}, {Cucciati}, {Mellier}, {Moreau}, \&
  {Paioro}}]{vvds_wide}
{Garilli}, B., {Le F{\`e}vre}, O., {Guzzo}, L., {et~al.} 2008, \aap, 486, 683

\bibitem[{{Geller} \& {Huchra}(1989)}]{Geller1988CfA2}
{Geller}, M.~J. \& {Huchra}, J.~P. 1989, Science, 246, 897

\bibitem[{{Giovanelli} {et~al.}(1986){Giovanelli}, {Haynes}, \&
  {Chincarini}}]{GiovanelliHaynes1986}
{Giovanelli}, R., {Haynes}, M.~P., \& {Chincarini}, G.~L. 1986, \apj, 300, 77

\bibitem[{{Goranova} {et~al.}(2009){Goranova}, {Hudelot}, {Magnard},
  {McCracken}, {Mellier}, {Monnerville}, {Schultheis}, {Semah}, {Cuillandre},
  \& {Aussel}}]{CFHTLS}
{Goranova}, Y., {Hudelot}, P., {Magnard}, F., {et~al.} 2009,
  http://terapix.iap.fr/cplt/table\_syn\_T0006.html

\bibitem[{{Grogin} {et~al.}(2011){Grogin}, {Kocevski}, {Faber}, {Ferguson},
  {Koekemoer}, {Riess}, {Acquaviva}, {Alexander}, {Almaini}, {Ashby}, {Barden},
  {Bell}, {Bournaud}, {Brown}, {Caputi}, {Casertano}, {Cassata}, {Castellano},
  {Challis}, {Chary}, {Cheung}, {Cirasuolo}, {Conselice}, {Roshan Cooray},
  {Croton}, {Daddi}, {Dahlen}, {Dav{\'e}}, {de Mello}, {Dekel}, {Dickinson},
  {Dolch}, {Donley}, {Dunlop}, {Dutton}, {Elbaz}, {Fazio}, {Filippenko},
  {Finkelstein}, {Fontana}, {Gardner}, {Garnavich}, {Gawiser}, {Giavalisco},
  {Grazian}, {Guo}, {Hathi}, {H{\"a}ussler}, {Hopkins}, {Huang}, {Huang},
  {Jha}, {Kartaltepe}, {Kirshner}, {Koo}, {Lai}, {Lee}, {Li}, {Lotz}, {Lucas},
  {Madau}, {McCarthy}, {McGrath}, {McIntosh}, {McLure}, {Mobasher},
  {Moustakas}, {Mozena}, {Nandra}, {Newman}, {Niemi}, {Noeske}, {Papovich},
  {Pentericci}, {Pope}, {Primack}, {Rajan}, {Ravindranath}, {Reddy}, {Renzini},
  {Rix}, {Robaina}, {Rodney}, {Rosario}, {Rosati}, {Salimbeni}, {Scarlata},
  {Siana}, {Simard}, {Smidt}, {Somerville}, {Spinrad}, {Straughn}, {Strolger},
  {Telford}, {Teplitz}, {Trump}, {van der Wel}, {Villforth}, {Wechsler},
  {Weiner}, {Wiklind}, {Wild}, {Wilson}, {Wuyts}, {Yan}, \& {Yun}}]{candels}
{Grogin}, N.~A., {Kocevski}, D.~D., {Faber}, S.~M., {et~al.} 2011, \apjs, 197,
  35

\bibitem[{{Grosb{\o}l} {et~al.}(2005){Grosb{\o}l}, {Banse}, {Tody}, {Cotton},
  {Cornwell}, {Ponz}, {Ignatius}, {Linde}, {van der Hulst}, {Burwitz},
  {Giaretta}, {Pasian}, {Garilli}, {Pence}, \& {Shaw}}]{fase1}
{Grosb{\o}l}, P., {Banse}, K., {Tody}, D., {et~al.} 2005, in Astronomical
  Society of the Pacific Conference Series, Vol. 347, Astronomical Data
  Analysis Software and Systems XIV, ed. P.~{Shopbell}, M.~{Britton}, \&
  R.~{Ebert}, 124

\bibitem[{{Hammersley} {et~al.}(2010){Hammersley}, {Christensen}, {Dekker},
  {Izzo}, {Selman}, {Bristow}, {Bourget}, {Castillo}, {Downing}, {Haddad},
  {Hilker}, {Lizon}, {Lucuix}, {Mainieri}, {Mieske}, {Reinero}, {Rejkuba},
  {Rojas}, {Smette}, {Urrutia Del Rio}, {Valenzuela}, \&
  {Wolff}}]{flexure_comp}
{Hammersley}, P., {Christensen}, L., {Dekker}, H., {et~al.} 2010, The
  Messenger, 142, 8

\bibitem[{{Hill} \& {Salinari}(1998)}]{lbt}
{Hill}, J.~M. \& {Salinari}, P. 1998, in Society of Photo-Optical
  Instrumentation Engineers (SPIE) Conference Series, Vol. 3352, Society of
  Photo-Optical Instrumentation Engineers (SPIE) Conference Series, ed. L.~M.
  {Stepp}, 23--33

\bibitem[{{Karampelas} {et~al.}(2012){Karampelas}, {Kontizas},
  {Rocca-Volmerange}, {Bellas-Velidis}, {Kontizas}, {Livanou}, {Tsalmantza}, \&
  {Dapergolas}}]{gaia2}
{Karampelas}, A., {Kontizas}, M., {Rocca-Volmerange}, B., {et~al.} 2012, \aap,
  538, A38

\bibitem[{{Kontizas} {et~al.}(2011){Kontizas}, {Bellas-Velidis},
  {Rocca-Volmerange}, {Kontizas}, {Tsalmantza}, {Livanou}, {Dapergolas}, \&
  {Karampelas}}]{gaia1}
{Kontizas}, M., {Bellas-Velidis}, I., {Rocca-Volmerange}, B., {et~al.} 2011, in
  EAS Publications Series, Vol.~45, EAS Publications Series, 337--342

\bibitem[{{Laureijs} {et~al.}(2011){Laureijs}, {Amiaux}, {Arduini},
  {Augu{\`e}res}, {Brinchmann}, {Cole}, {Cropper}, {Dabin}, {Duvet}, {Ealet},
  \& et~al.}]{euclidRB}
{Laureijs}, R., {Amiaux}, J., {Arduini}, S., {et~al.} 2011, ArXiv e-prints
  astro-ph/1110.3193

\bibitem[{{Le Fevre} {et~al.}(1995){Le Fevre}, {Crampton}, {Lilly}, {Hammer},
  \& {Tresse}}]{CFRS}
{Le Fevre}, O., {Crampton}, D., {Lilly}, S.~J., {Hammer}, F., \& {Tresse}, L.
  1995, \apj, 455, 60

\bibitem[{{Le F{\`e}vre} {et~al.}(2002){Le F{\`e}vre}, {Mancini}, {Saisse},
  {Brau-Nogu{\'e}}, {Caputi}, {Castinel}, {D'Odorico}, {Garilli}, {Kissler},
  {Lucuix}, {Mancini}, {Pauget}, {Sciarretta}, {Scodeggio}, {Tresse},
  {Maccagni}, {Picat}, \& {Vettolani}}]{instrument2}
{Le F{\`e}vre}, O., {Mancini}, D., {Saisse}, M., {et~al.} 2002, The Messenger,
  109, 21

\bibitem[{{Le F{\`e}vre} {et~al.}(2000){Le F{\`e}vre}, {Saisse}, {Mancini},
  {Vettolani}, {Maccagni}, {Picat}, {Mellier}, {Mazure}, {Cuby}, {Delabre},
  {Garilli}, {Hill}, {Prieto}, {Voet}, {Arnold}, {Brau-Nogue}, {Cascone},
  {Conconi}, {Finger}, {Huster}, {Laloge}, {Lucuix}, {Mattaini}, {Schipani},
  {Waultier}, {Zerbi}, {Avila}, {Beletic}, {D'Odorico}, {Moorwood}, {Monnet},
  \& {Reyes Moreno}}]{instrument1}
{Le F{\`e}vre}, O., {Saisse}, M., {Mancini}, D., {et~al.} 2000, in Society of
  Photo-Optical Instrumentation Engineers (SPIE) Conference Series, Vol. 4008,
  Society of Photo-Optical Instrumentation Engineers (SPIE) Conference Series,
  ed. M.~{Iye} \& A.~F. {Moorwood}, 546--557

\bibitem[{{Le F{\`e}vre} {et~al.}(2005){Le F{\`e}vre}, {Vettolani}, {Garilli},
  {Tresse}, {Bottini}, {Le Brun}, {Maccagni}, {Picat}, {Scaramella},
  {Scodeggio}, {Zanichelli}, {Adami}, {Arnaboldi}, {Arnouts}, {Bardelli},
  {Bolzonella}, {Cappi}, {Charlot}, {Ciliegi}, {Contini}, {Foucaud},
  {Franzetti}, {Gavignaud}, {Guzzo}, {Ilbert}, {Iovino}, {McCracken}, {Marano},
  {Marinoni}, {Mathez}, {Mazure}, {Meneux}, {Merighi}, {Paltani}, {Pell{\`o}},
  {Pollo}, {Pozzetti}, {Radovich}, {Zamorani}, {Zucca}, {Bondi}, {Bongiorno},
  {Busarello}, {Lamareille}, {Mellier}, {Merluzzi}, {Ripepi}, \&
  {Rizzo}}]{vvds_main}
{Le F{\`e}vre}, O., {Vettolani}, G., {Garilli}, B., {et~al.} 2005, \aap, 439,
  845

\bibitem[{{Lilly} {et~al.}(2007){Lilly}, {Le F{\`e}vre}, {Renzini}, {Zamorani},
  {Scodeggio}, {Contini}, {Carollo}, {Hasinger}, {Kneib}, {Iovino}, {Le Brun},
  {Maier}, {Mainieri}, {Mignoli}, {Silverman}, {Tasca}, {Bolzonella},
  {Bongiorno}, {Bottini}, {Capak}, {Caputi}, {Cimatti}, {Cucciati}, {Daddi},
  {Feldmann}, {Franzetti}, {Garilli}, {Guzzo}, {Ilbert}, {Kampczyk}, {Kovac},
  {Lamareille}, {Leauthaud}, {Borgne}, {McCracken}, {Marinoni}, {Pello},
  {Ricciardelli}, {Scarlata}, {Vergani}, {Sanders}, {Schinnerer}, {Scoville},
  {Taniguchi}, {Arnouts}, {Aussel}, {Bardelli}, {Brusa}, {Cappi}, {Ciliegi},
  {Finoguenov}, {Foucaud}, {Franceschini}, {Halliday}, {Impey}, {Knobel},
  {Koekemoer}, {Kurk}, {Maccagni}, {Maddox}, {Marano}, {Marconi}, {Meneux},
  {Mobasher}, {Moreau}, {Peacock}, {Porciani}, {Pozzetti}, {Scaramella},
  {Schiminovich}, {Shopbell}, {Smail}, {Thompson}, {Tresse}, {Vettolani},
  {Zanichelli}, \& {Zucca}}]{zcosmos_main}
{Lilly}, S.~J., {Le F{\`e}vre}, O., {Renzini}, A., {et~al.} 2007, \apjs, 172,
  70

\bibitem[{{Lupton} {et~al.}(2002){Lupton}, {Ivezic}, {Gunn}, {Knapp},
  {Strauss}, \& {Yasuda}}]{Lupton2002}
{Lupton}, R.~H., {Ivezic}, Z., {Gunn}, J.~E., {et~al.} 2002, in Society of
  Photo-Optical Instrumentation Engineers (SPIE) Conference Series, Vol. 4836,
  Society of Photo-Optical Instrumentation Engineers (SPIE) Conference Series,
  ed. J.~A. {Tyson} \& S.~{Wolff}, 350--356

\bibitem[{{Magrini} {et~al.}(2012){Magrini}, {Sommariva}, {Cresci}, {Sani},
  {Galametz}, {Mannucci}, {Petropoulou}, \& {Fumana}}]{luciReduction}
{Magrini}, L., {Sommariva}, V., {Cresci}, G., {et~al.} 2012, ArXiv e-prints

\bibitem[{{Mandel} {et~al.}(2000){Mandel}, {Appenzeller}, {Bomans},
  {Eisenhauer}, {Grimm}, {Herbst}, {Hofmann}, {Lehmitz}, {Lemke}, {Lehnert},
  {Lenzen}, {Luks}, {Mohr}, {Seifert}, {Thatte}, {Weiser}, \& {Xu}}]{lucifer}
{Mandel}, H., {Appenzeller}, I., {Bomans}, D., {et~al.} 2000, in Society of
  Photo-Optical Instrumentation Engineers (SPIE) Conference Series, Vol. 4008,
  Society of Photo-Optical Instrumentation Engineers (SPIE) Conference Series,
  ed. M.~{Iye} \& A.~F. {Moorwood}, 767--777

\bibitem[{{Marchetti} {et~al.}(2012){Marchetti}, {Granett}, {Guzzo}, {Fritz},
  {Garilli}, {Scodeggio}, {Abbas}, {Adami}, {Arnouts}, {Bolzonella}, {Bottini},
  {Cappi}, {Coupon}, {Cucciati}, {De Lucia}, {de la Torre}, {Franzetti},
  {Fumana}, {Ilbert}, {Iovino}, {Krywult}, {Le Brun}, {Le Fevre}, {Maccagni},
  {Malek}, {Marulli}, {McCracken}, {Meneux}, {Paioro}, {Polletta}, {Pollo},
  {Schlagenhaufer}, {Tasca}, {Tojeiro}, {Vergani}, {Zanichelli}, {Bel},
  {Bersanelli}, {Blaizot}, {Branchini}, {Burden}, {Davidzon}, {Di Porto},
  {Guennou}, {Marinoni}, {Mellier}, {Moscardini}, {Nichol}, {Peacock},
  {Percival}, {Phleps}, {Schimd}, {Wolk}, \& {Zamorani}}]{Marchetti2012}
{Marchetti}, A., {Granett}, B.~R., {Guzzo}, L., {et~al.} 2012, ArXiv e-prints
  astro-ph/1207.4374

\bibitem[{{Nastasi} {et~al.}(2012){Nastasi}, {Fassbender}, {Bohringer},
  {Pierini}, {Verdugo}, {Garilli}, \& {Franzetti}}]{fvipgi}
{Nastasi}, A.and~{Scodeggio}, M., {Fassbender}, R., {Bohringer}, H., {et~al.}
  2012, \aap, accepted for publication

\bibitem[{{Paioro} {et~al.}(2008){Paioro}, {Chiappetti}, {Garilli},
  {Franzetti}, {Fumana}, \& {Scodeggio}}]{dart}
{Paioro}, L., {Chiappetti}, L., {Garilli}, B., {et~al.} 2008, in Astronomical
  Society of the Pacific Conference Series, Vol. 394, Astronomical Data
  Analysis Software and Systems XVII, ed. R.~W. {Argyle}, P.~S. {Bunclark}, \&
  J.~R. {Lewis}, 397

\bibitem[{{Paioro} {et~al.}(2010){Paioro}, {Garilli}, {Grosb{\o}l}, {Tody},
  {Fenouillet}, {Granet}, \& {Surace}}]{fase2}
{Paioro}, L., {Garilli}, B., {Grosb{\o}l}, P., {et~al.} 2010, in Astronomical
  Society of the Pacific Conference Series, Vol. 434, Astronomical Data
  Analysis Software and Systems XIX, ed. {Y.~Mizumoto, K.-I.~Morita, \&
  M.~Ohishi}, 349

\bibitem[{{Pogge} {et~al.}(2010){Pogge}, {Atwood}, {Brewer}, {Byard},
  {Derwent}, {Gonzalez}, {Martini}, {Mason}, {O'Brien}, {Osmer}, {Pappalardo},
  {Steinbrecher}, {Teiga}, \& {Zhelem}}]{mods}
{Pogge}, R.~W., {Atwood}, B., {Brewer}, D.~F., {et~al.} 2010, in Society of
  Photo-Optical Instrumentation Engineers (SPIE) Conference Series, Vol. 7735,
  Society of Photo-Optical Instrumentation Engineers (SPIE) Conference Series

\bibitem[{{Schlegel} {et~al.}(2007){Schlegel}, {Blanton}, {Eisenstein},
  {Gillespie}, {Gunn}, {Harding}, {McDonald}, {Nichol}, {Padmanabhan},
  {Percival}, {Richards}, {Rockosi}, {Roe}, {Ross}, {Schneider}, {Strauss},
  {Weinberg}, \& {White}}]{BOSS}
{Schlegel}, D.~J., {Blanton}, M., {Eisenstein}, D., {et~al.} 2007, in Bulletin
  of the American Astronomical Society, Vol.~39, American Astronomical Society
  Meeting Abstracts, 132.29

\bibitem[{{Scodeggio} {et~al.}(2009){Scodeggio}, {Franzetti}, {Garilli}, {Le
  F{\`e}vre}, \& {Guzzo}}]{shortSlit}
{Scodeggio}, M., {Franzetti}, P., {Garilli}, B., {Le F{\`e}vre}, O., \&
  {Guzzo}, L. 2009, The Messenger, 135, 13

\bibitem[{{Scodeggio} {et~al.}(2005){Scodeggio}, {Franzetti}, {Garilli},
  {Zanichelli}, {Paltani}, {Maccagni}, {Bottini}, {Le Brun}, {Contini},
  {Scaramella}, {Adami}, {Bardelli}, {Zucca}, {Tresse}, {Ilbert}, {Foucaud},
  {Iovino}, {Merighi}, {Zamorani}, {Gavignaud}, {Rizzo}, {McCracken}, {Le
  F{\`e}vre}, {Picat}, {Vettolani}, {Arnaboldi}, {Arnouts}, {Bolzonella},
  {Cappi}, {Charlot}, {Ciliegi}, {Guzzo}, {Marano}, {Marinoni}, {Mathez},
  {Mazure}, {Meneux}, {Pell{\`o}}, {Pollo}, {Pozzetti}, \& {Radovich}}]{vipgi}
{Scodeggio}, M., {Franzetti}, P., {Garilli}, B., {et~al.} 2005, \pasp, 117,
  1284

\bibitem[{{Shectman} {et~al.}(1996){Shectman}, {Landy}, {Oemler}, {Tucker},
  {Lin}, {Kirshner}, \& {Schechter}}]{LasCampanas}
{Shectman}, S.~A., {Landy}, S.~D., {Oemler}, A., {et~al.} 1996, \apj, 470, 172

\bibitem[{{Szalay} {et~al.}(2002){Szalay}, {Gray}, \&
  {VandenBerg}}]{Szalay2002}
{Szalay}, A.~S., {Gray}, J., \& {VandenBerg}, J. 2002, in Society of
  Photo-Optical Instrumentation Engineers (SPIE) Conference Series, Vol. 4836,
  Society of Photo-Optical Instrumentation Engineers (SPIE) Conference Series,
  ed. J.~A. {Tyson} \& S.~{Wolff}, 333--338

\bibitem[{{Vettolani} {et~al.}(1997){Vettolani}, {Zucca}, {Zamorani}, {Cappi},
  {Merighi}, {Mignoli}, {Stirpe}, {MacGillivray}, {Collins}, {Balkowski},
  {Cayatte}, {Maurogordato}, {Proust}, {Chincarini}, {Guzzo}, {Maccagni},
  {Scaramella}, {Blanchard}, \& {Ramella}}]{ESP}
{Vettolani}, G., {Zucca}, E., {Zamorani}, G., {et~al.} 1997, \aap, 325, 954

\bibitem[{{Warren} {et~al.}(2007){Warren}, {Hambly}, {Dye}, {Almaini}, {Cross},
  {Edge}, {Foucaud}, {Hewett}, {Hodgkin}, {Irwin}, {Jameson}, {Lawrence},
  {Lucas}, {Adamson}, {Bandyopadhyay}, {Bryant}, {Collins}, {Davis}, {Dunlop},
  {Emerson}, {Evans}, {Gonzales-Solares}, {Hirst}, {Jarvis}, {Kendall}, {Kerr},
  {Leggett}, {Lewis}, {Mann}, {McLure}, {McMahon}, {Mortlock}, {Rawlings},
  {Read}, {Riello}, {Simpson}, {Smith}, {Sutorius}, {Targett}, \&
  {Varricatt}}]{ukidss}
{Warren}, S.~J., {Hambly}, N.~C., {Dye}, S., {et~al.} 2007, \mnras, 375, 213

\end{thebibliography}
\clearpage

\begin{deluxetable}{ccccc}
\tablecolumns{4}
\tablewidth{0pc}
\tablecaption{Easylife performance on data reduction
\label{reduction_stats}}
\tablehead{
\colhead
{Failure reason}& {failure rate}  }
\startdata
Spectra location  & 2\%  \\
Wavelength calibration    & 0.5\% \\
Target detection    & 5.5\%   \\
Total               & 8\%  \\
\enddata
\end{deluxetable}

\begin{deluxetable}{ccccc}
\tablecolumns{2}
\tablewidth{0pc}
\tablecaption{Automatic redshift measurement performance 
\label{perfTable}}
\tablehead{
\colhead
{EZ flag}& total spectra & correct redshifts & success rate 
}
\startdata
any  &35903  & 27322 &76\%   \\
3-4  & 20043 & 18889 & 94\%     \\
2    & 2213  & 1677  & 76\%   \\
9    & 1548  & 1188  & 77\%   \\
1    & 2790  & 1970  & 71\%    \\
0    & 6941  & 3598  & 52\%   \\
\enddata
\end{deluxetable}

\begin{deluxetable}{cccc}
\tablecolumns{4}
\tablewidth{0pc}
\tablecaption{Automatic and humanly assigned flags comparison
\label{real_flag_table}}
\tablehead{
\colhead
{} & \multicolumn{3}{c}{human flag}\\
\cline{2-4}
{EZ flag} & {3-4} & {2-9} & {1} }
\startdata
3-4 & 81 \%& 16 \%& 3 \%  \\ 
2-9 & 32 \%& 55 \%& 13 \%  \\ 
1   & 40 \%& 41 \%& 19 \%  \\ 
0   & 27 \%& 48 \%& 25 \%  \\ 
\enddata
\end{deluxetable}

\clearpage
\begin{figure*}
  \includegraphics[clip=true,scale=0.7]{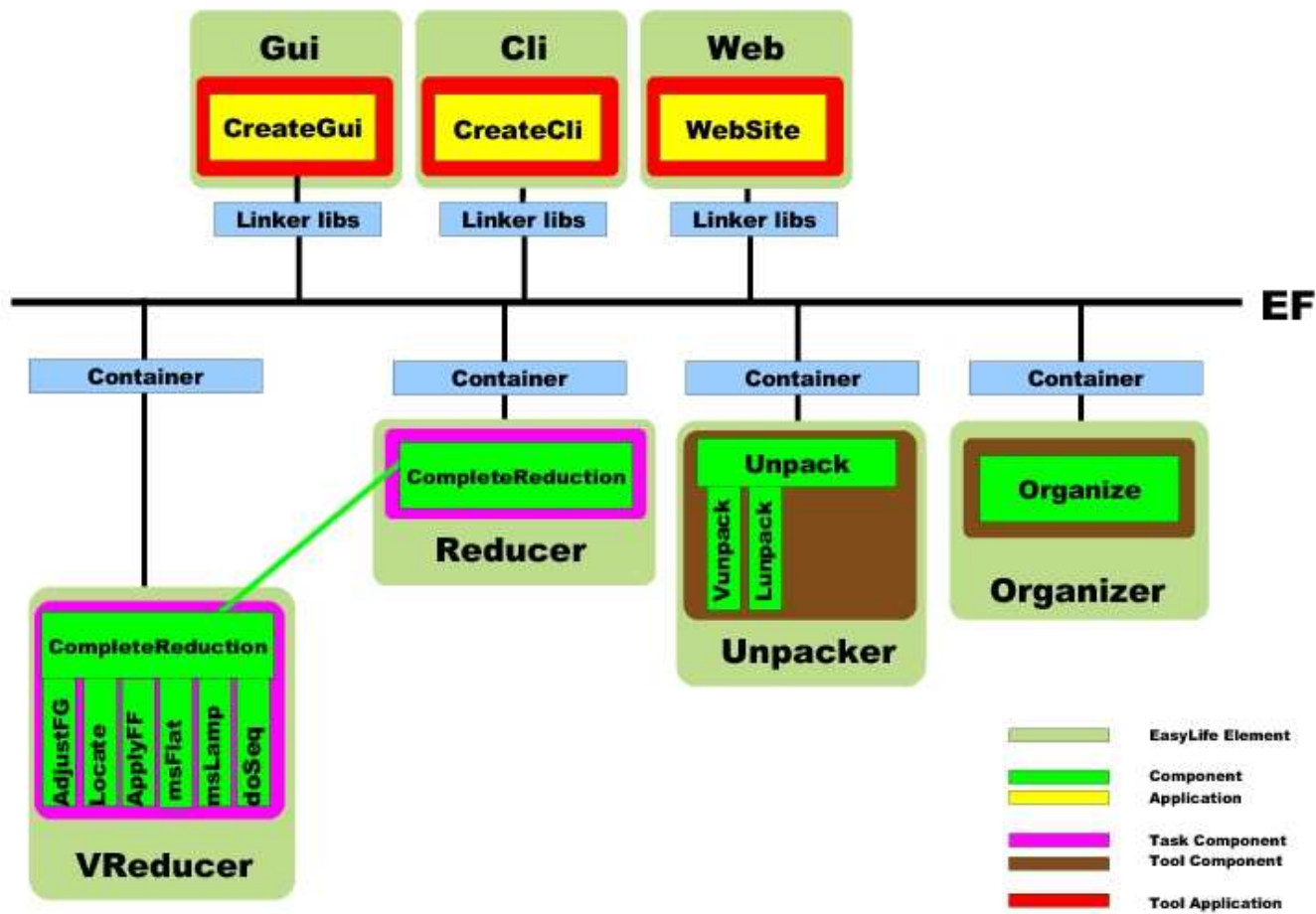}
  \caption{FASE architecture as implemented for the {\it Easylife} system. On the
    top part, we find the application and presentation layer. The bottom part
    shows the three main containers (Reducer, Unpacker and Organizer) with
    their respective components. Everything is
           linked together by the execution framework (EF) provided by an
           early prototype of FASE environment.}
  \label{design}
\end{figure*}

\end{document}